\documentclass[10pt,letterpaper]{article}
\usepackage[top=0.85in,left=2.75in,footskip=0.75in]{geometry}

\usepackage{amsmath,amssymb}

\usepackage{changepage}

\usepackage[utf8x]{inputenc}

\usepackage{textcomp,marvosym}

\usepackage{cite}

\usepackage{nameref,hyperref}

\usepackage{microtype}
\DisableLigatures[f]{encoding = *, family = * }

\usepackage[table]{xcolor}

\usepackage{array}
\usepackage{booktabs}
\usepackage{multirow}
\newcolumntype{+}{!{\vrule width 2pt}}
\newcolumntype{L}[1]{>{\raggedright\let\newline\\\arraybackslash\hspace{0pt}}m{#1}}
\newcolumntype{C}[1]{>{\centering\let\newline\\\arraybackslash\hspace{0pt}}m{#1}}
\newcolumntype{R}[1]{>{\raggedleft\let\newline\\\arraybackslash\hspace{0pt}}m{#1}}
\newlength\savedwidth

\raggedright
\usepackage[aboveskip=1pt,labelfont=bf,labelsep=period,justification=raggedright,singlelinecheck=off]{caption}

\makeatletter
\renewcommand{\@biblabel}[1]{\quad#1.}
\makeatother

\date{}

\usepackage{lastpage,fancyhdr,graphicx}
\usepackage{epstopdf}
\usepackage{nameref}
\AppendGraphicsExtensions{.tif}
\fancyheadoffset[L]{2.25in}
\fancyfootoffset[L]{2.25in}

\begin{document}
\vspace*{0.2in}

\begin{flushleft}
{\Large
\textbf\newline{Mapping Three-dimensional Urban Structure by Fusing Landsat and Global Elevation Data} 
}
\newline
\\
Panshi Wang\textsuperscript{1,*},
Chengquan Huang\textsuperscript{1},
James C. Tilton\textsuperscript{2}
\\
\bigskip
\textbf{1} Department of Geographical Sciences, University of Maryland, College Park, Maryland, USA\\
\textbf{2} Computational and Information Sciences and Technology Office, NASA Goddard Space Flight Center, Greenbelt, Maryland, USA
\\
\bigskip

%
%



* pswang@umd.edu

\end{flushleft}
\section*{Abstract}
To meet the challenges of global urbanization, earth observation information is greatly needed. The lack of global three-dimensional (3D) urban structure data has been a major limiting factor in important urban applications such as population mapping, disaster vulnerability assessment, and climate change adaptation. Due to limited data availability, remote sensing data haven been mainly used to characterize 3D urban structure at the city scale. In this study, we propose a method to map 3D urban structure using freely available Landsat and global elevation data. Building on an object-based machine learning approach, the synergy of Landsat and elevation data were used to estimate building height and volume at 30 m resolution. This method has been tested for the entire country of England and yielded a root-mean-square error (RMSE) of 1.61 m for building height and an RMSE of 1,142 m\textsuperscript{3} for building volume. The results in this study represent the first attempt to use open data for urban structure characterization at the country scale. Our results demonstrated the utility of these data for large-scale urban studies and the potential of generating global 3D urban structure data products using a fusion-based approach.

\section*{Introduction}
Urbanization has become a global phenomenon. More than half of the world’s population now dwells in urban areas and the urban population is expected to reach two thirds of the world’s population by 2050 \cite{un2015world}. For the last three decades of the 20\textsuperscript{th} century, the rate of urban land expansion has been even higher than the rate of urban population growth, a trend that is expected to continue in the future \cite{seto2011meta}. Both the environment and societies face great pressures associated with rapid urbanization, including the urban heat island effect \cite{arnfield2003two}, air and water quality degradation \cite{duncan2016space,arnold1996impervious}, agricultural land loss\cite{seto2016hidden}, and increasing demands for essential infrastructure (e.g., water supply) \cite{larsen2016emerging}. In response to these problems, a growing body of urban research have been established to monitor, understand and model the trends and impacts of global urbanization \cite{wang2012global,seto2009global,seto2012global}. Remote sensing has been proven instrumental to achieving these goals, particulary by providing increasingly detailed information on the spatial extent and form of urban areas \cite{wentz2014supporting}. Recently, several fine-resolution global urban data products \cite{esch2011characterization,esch2013urban,pesaresi2016ghs,GMIS,HBASE} have been made available. These data products lay the foundation for a more comprehensive urban science and better solutions for sustainable urbanization. 

However, current urban data lack a good representation of three-dimensional (3D) urban structure, which has been a missing element of our understanding of urban landscape and growth trajectory at the global scale \cite{frolking2013global}. Urbanization, as a Land Cover/Land Use (LCLU) change process, is not only manifested through two-dimensional (2D) urban expansion and intensification. Rather, vertical growth has also been an important characteristic of urbanization as demonstrated by studies across the globe \cite{frolking2013global,salvati2013changes,magarotto2016vertical,henderson2016building,zhang2017detecting}. Indeed, the 3D structure of an urban area is closely linked to its physical properties as well as the spatial distribution of its population and human activities \cite{souch2006applied,gong2011icesat,masson2006urban}. First, inclusion of urban structural information has been found important for modeling climatic impacts of urbanization (e.g., the urban heat island effect and impacts of urbanization on precipitation patterns) \cite{arnfield2003two,jin2005inclusion,masson2006urban,souch2006applied}. Second, a reliable, spatially detailed and updatable database of urban structural information is needed for key applications such as disaster vulnerability assessment \cite{geiss2016estimation} and population distribution mapping \cite{lu2011volumetric,qiu2010spatial}. Finally, urban structure is also related to many urban research topics with policy implications, such as finding energy-efficient urban layout and redeveloping urban slums \cite{seto2013remote,ratti2005energy,henderson2016building,RePEc:ehl:lserod:66536}. Therefore, without proper characterization of urbanization in the 3D space, it is impossible to fully understand the functions and mechanisms of the urban system, its interactions with the environment, and the opportunities and challenges it will present to humanity.

A growing amount of interest has shifted from 2D properties of urban landscape to the vertical dimension within the urban remote sensing community. To date, the bulk of the literature on the remote sensing of 3D urban structure relies on the light detection and ranging (LIDAR) technology, particularly at the scale of a city \cite{cheng2011trend,gong2011icesat,gonzalez2013automated,zheng2015model}. Optical image stereography \cite{sirmacek2012performance,wurm2014investigating} and interferometric synthetic aperture radar (InSAR; \cite{gamba2000detection,thiele2010combining}) have also been proven to be effective technologies to map urban structural parameters such as building height (BH) and building volume (BV). While lidar provides an unmatched vertical accuracy, its applications are often limited by incomplete spatial coverage due to its sampling nature. Although to a lesser extent, spatial coverage limitations also apply to high-resolution stereo optical images and sometimes also InSAR images \cite{lu2007interferometric,wurm2014investigating}. 

Even for areas with satisfactory data coverage, there are two major technical challenges for urban 3D structure mapping. First, all three technological approaches typically measure the top of surface height, also known as the digital surface model (DSM). In the case of buildings, DSMs provide the height of buildings plus the ground height (i.e., the digital terrain mode [DTM]). To extract BH/BV, either direct approaches that separate the ground and building signals \cite{gamba2000detection,soergel2003iterative} or indirect approaches that remove the ground height from DSMs need to be employed \cite{geiss2015normalization,meng2012detect}. However, direct approaches rely on the modeling of InSAR or lidar signals from individual buildings and therefore works best with high resolution data, while the indirect approaches adopt filtering methods to substract the ground height from DSMs, which also requires a sufficient resolution in densely urbanized areas. Second, buildings are not the only vertical structures on the ground surface. Trees, for example, must be removed for BH/BV mapping. To address this issue, direct and indirect approaches have also been applied to high resolution data\cite{gamba2000detection,zheng2015model}. In addition to these direct and indirect approaches, building boundaries derived from vector data layers or optical imagery classifications have also been used as ancillary information to address these two technical issues \cite{thiele2010combining,wegner2014combining}.

Despite the success of these existing methods at small scales, data constraints remain the biggest challenge for mapping large-scale urban structure. Since the Shuttle Radar Topography Mission (SRTM) \cite{van2001shuttle}, several satellite missions have generated global DSM data products \cite{tachikawa2011characteristics,tadono2016generation,krieger2007tandem}. To date, medium-resolution DSMs are the best freely available sources of large-scale height information, raising the question if they could be used for urban structure characterization, a possibility first envisioned by \cite{nghiem2001global}. Unfortunately, due to insufficient resolutions of freely available global DSMs, existing methods may not be directly applicable.

In this study, the fusion of Landsat and global DSMs is investigated to address the challenges of large-scale urban structure mapping. Our objectives are to develop a method for mapping BH/BV at 30 m resolution, to demonstrate the effectiveness of our method for a large area, and to explore the usefulness of 3D urban mapping in socioeconomic studies. By leveraging the spatial information from Landsat-based segmentation of urban land patches as demonstrated in the our previous study \cite{hotex}, we propose using a suite of object-based height metrics derived from global DSMs and machine learning algorithms to estimate BH/BV. This method is applied to the entire country of England to produce 30 m BH/BV maps. To the best of the authors' knowledge, this study is the first attempt to produce wall-to-wall maps of BH/BV for such a large area.

\section*{Study Area}
\begin{figure}[!h]
\caption{{\bf The extent of the study area.}
The study area includes the entire country of England (red polygon). The gray lines illustrate the boundaries of the WRS-2 Landsat tiles covering the study area. Major metropolitan areas in England defined by the Organization for Economic Co-operation and Development (OECD; \cite{brezzi2012redefining}) are shown as polygons filled with gray color.}
\label{fig1}
\includegraphics[width=5.2in]{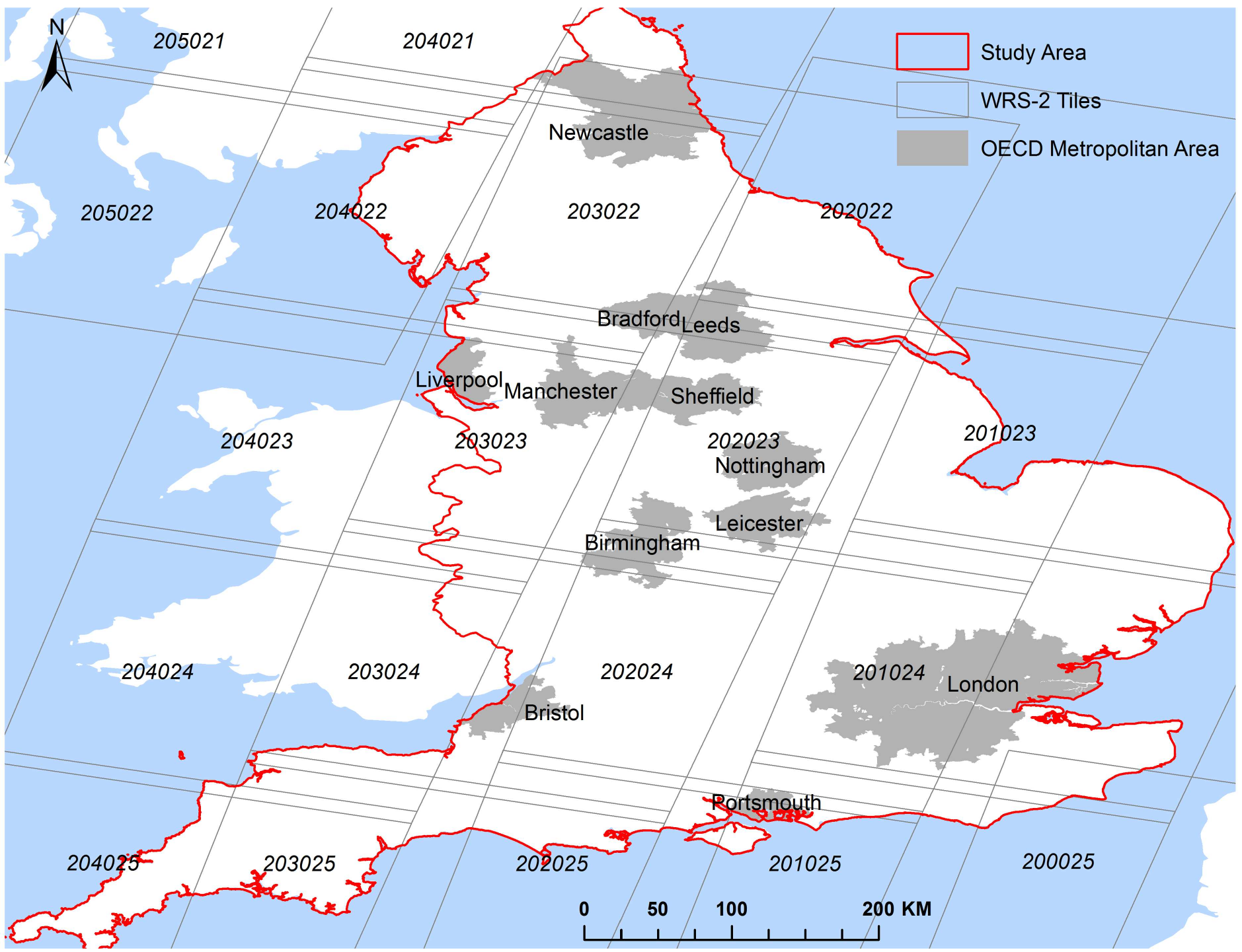}
\end{figure}

The study area of this paper is the entire country of England (see Fig~\ref{fig1}), covering 130,279 km\textsuperscript{2} of land area and with a population of 53,012,456 (according to the 2011 census). Based on two land cover maps (LCMs) for 2000 and 2015, the total urban area of England is 13,788 km\textsuperscript{2} and 14,194 km\textsuperscript{2}, respectively, which yields a 0.2\% annual increase rate \cite{fuller2002countryside,lcm2015}. Also, according to the United Nations (United Nations 2015), the average annual rate of population growth for the United Kingdom (UK) is 0.45\%, 0.98\%, and 0.65\% for the time periods 2000–2005, 2005–2010, and 2010–2015, respectively, which is much lower than that of mid-income and low-income countries. Furthermore, from 2000 to 2012, the total business floor space of England moderately increased from 527,058 m\textsuperscript{2} to 544,414 m\textsuperscript{2}, with an annual growth rate of 0.3\% \cite{valuationoffice2012}, compared with an annual growth rate of 2\% of the total commercial floor space in the United States from 1999 to 2012 \cite{EIA2016}. Therefore, it was reasonable to assume that the urban growth rates in both the 2D space and the vertical dimension were relatively slow during the 1998–2016 period, although much faster urban growth might exist in certain areas including the suburban areas of large metropolises such as London.

The relative slow urban growth of England was an important factor in our study area selection. We used multi-source data sets for mapping BH/BV and for the training and validation of machine learning models (see the \nameref{sec:data} section for detailed descriptions of all used data). Limited by data acquisition frequency, these data sets were collected over a wide time span (1998-2016). This would lead to potential discrepancies among (a) training/validation data and machine learning model input variables and (b) different input variables. Choosing England as the study area, we greatly limited the impact of such discrepancies. Year 2010 was designated as the nominal mapping year, because it is roughly the midpoint of the temporal range of the best available data sets.

\section*{Methods}
\begin{figure}[!h]
\begin{adjustwidth}{-2.25in}{0in} 
\caption{{\bf A conceptual framework of the methodology of this study.}}
\label{fig2}
\includegraphics[width=7.5in]{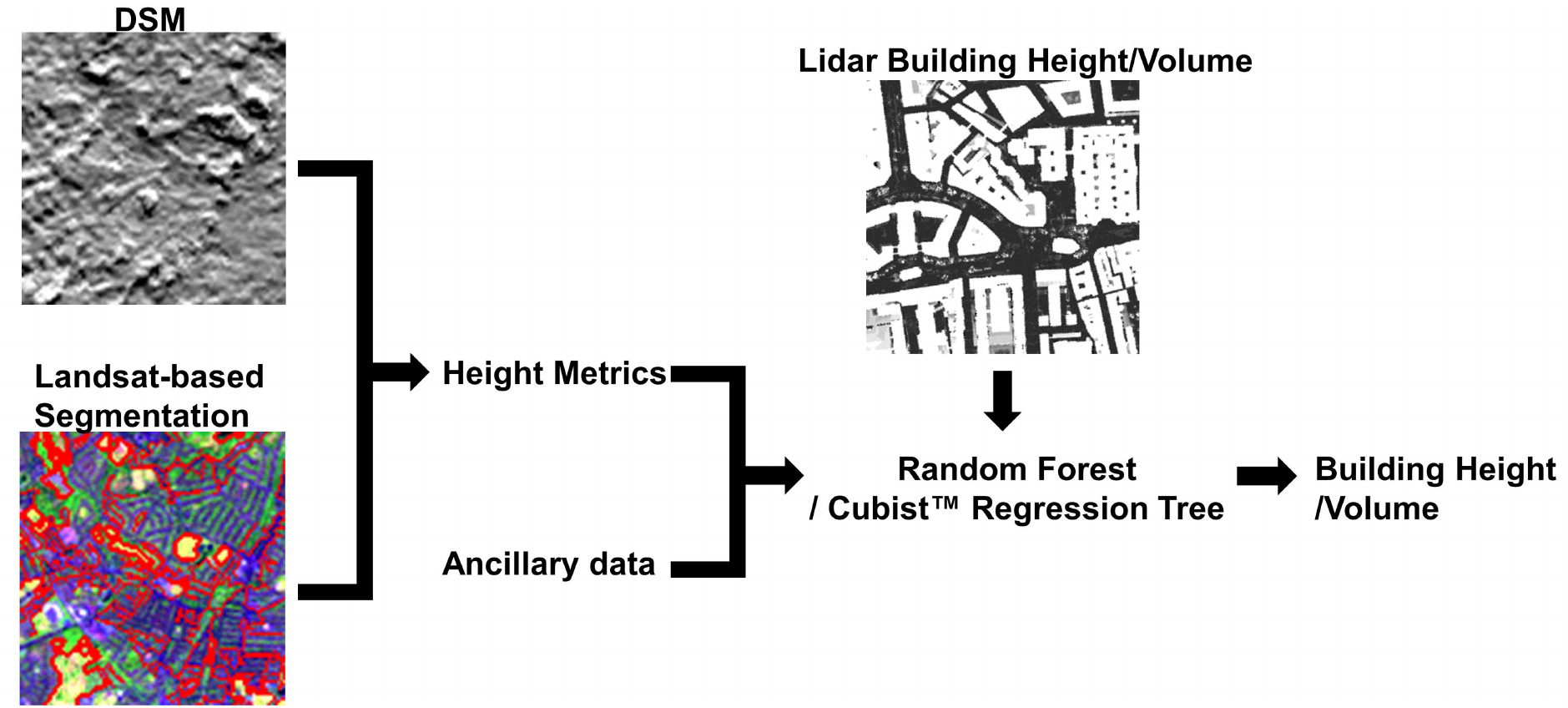}
\end{adjustwidth}
\end{figure}

Our method features the fusion of Landsat and DSM data using an object-based approach. A conceptual framework of the proposed method is provided in Fig \ref{fig2}. We describe the data used and individual steps taken in detail below.

\subsection*{Data}
\label{sec:data}
\subsubsection*{Lidar-based Building Height Data}
The UK Environment Agency (EA) has been collecting airborne lidar data since 1998. By 2016, about 75\% of England has been mapped at least once with accurate elevation measurements. The 1 m resolution DSM and DTM data sets based on lidar acquisitions during the 2006--2014 temporal period have been made freely available \cite{EA2016}. According to the EA, the absolute height error of the data set is within ±15 cm, while the relative height error is less than ±5 cm \cite{EA2016}, making it an excellent source of training and validation data for BH/BV estimation.

As described in the introduction, BH mapping with lidar data requires either accurate auxiliary building boundary information or the automatic separation of buildings and other vertical structures, such as trees. Here, we used an open database provided by Emu Analytics \cite{Emu2016}, which used building boundaries from the ordnance survey of Britain to separate heights of buildings and non-buildings. While the quality of this data set has not been fully assessed, it is the best reference data available and should provide sufficient accuracy for training and validation purposes.

\subsubsection*{Global Elevation Data}
Three global DSMs were used in this study including the Shuttle Radar Topography Mission (SRTM), the Advanced Spaceborne Thermal Emission and Reflection Radiometer (ASTER) Global Digital Elevation Model (GDEM), and the Advanced Land Observing Satellite (ALOS) World 3D–30 m (AW3D30) data sets. Table \ref{table1} lists the main characteristics of these DSMs. All three DSMs are posted at about 30 m spatial resolution and provide top of surface heights. Yet, according to validations at different locations, they perform differently in terms of vertical accuracy due to differences in data sources and processing methods. Although AW3D30 is less well validated given its recent release date, it is expected to have a better vertical accuracy than the other two DSMs because its data source, the Panchromatic Remote-sensing Instrument for Stereo Mapping (PRISM) sensor, has superior spatial resolution (2.5 m, \cite{tadono2016generation}). This is particularly important for urban applications because most urban features can be better resolved at a higher spatial resolution. One major shortcoming of AW3D30, however, is that it has data gaps as a result of cloud cover and gaps between satellite orbits \cite{tadono2016generation}. On the other hand, SRTM and ASTER GDEM have been widely utilized in different applications and assessed in different areas of the world \cite{mukherjee2013evaluation,satge2016absolute,small2015correlation}. Based on selected studies on the accuracy of these two DSMs, none of them consistently outperforms the other. The vertical accuracy varies among different terrain types. In urban areas, both SRTM and ASTER GDEM demonstrate an overestimation of the ground height and underestimation of the building top height \cite{small2015correlation}, while the ASTER GDEM may have a higher error in urban areas. This is because stereography in heterogeneous urban environments may require higher spatial resolution than what ASTER could provide. In summary, there is no clear best choice among these three DSMs considering both spatial coverage and vertical accuracy. Therefore, we included all three DSMs to map BH/BV assuming they have complimentary advantages. For the feature extraction processing steps, these DSMs were projected and resampled using a nearest neighbor resampling technique to match them with Landsat images.

\begin{table}[!ht]
\begin{adjustwidth}{-2.25in}{0in} 
\centering
\caption{
{\bf Characteristics of fine-resolution (30 m) global elevation datasets used in this study.}}
\begin{tabular}{C{2.5cm} C{2.5cm} C{3cm} C{2.5cm} C{6.5cm}}
  \toprule
  Name & Acquisition Year  & Data Source & Version & Vertical Accuracy (RMSE) \\
\midrule
\midrule \\
 SRTM & 2000 (11 days) & C/X band InSAR & SRTMGL1 v3 & 6.6 m \cite{satge2016absolute}, 5.53--12.77 m \cite{rodriguez2006global}, 17.76 m \cite{mukherjee2013evaluation}\\
\midrule \\
 ASTER GDEM & 2000--2011 & ASTER & v2 & 8.68 m \cite{tachikawa2011characteristics}, 9 m \cite{satge2016absolute}, 12.62 m \cite{mukherjee2013evaluation}\\
\midrule \\
 AW3D30 & 2006--2011  & PRISM & v2 & 4.4 m \cite{tadono2016generation}\\
  \bottomrule
\end{tabular}
\label{table1}
\end{adjustwidth}
\end{table}

\subsubsection*{Landsat Data}
We manually selected 24 Landsat 5 (L5) and Landsat 8 (L8) scenes covering the entire land mass of England. In the selection of these images, acquisitions close to the temporal range of DSM data sets (2000-2011) were preferred, because Landsat images need to be overlaid with DSM data sets for height information extraction. However, Landsat 7 images in the study area were found to be severely affected by the failure of the scan line corrector (SLC) \cite{arvidson2006landsat}. Also, cloud-free Landsat observation is scarce for many parts of England. Considering all these factors, 16 of the selected images were L5 images within the 2009–2011 temporal window, 5 were L5 images acquired during 2003–2006, and 3 were Landsat 8 images acquired between 2013 and 2014. All the selected scenes have been atmospherically corrected and converted to surface reflectance, which were available through the U.S. Geological Survey (USGS) Earth Resources Observation and Science (EROS) Center Science Processing Architecture (ESPA) on demand ordering service.

\subsubsection*{Landsat-based Urban Extent and Impervious Surface Data}
A global circa-2010 data set of impervious surface (IS) fraction have been produced by the Global Man-made Impervious Surface (GMIS) project \cite{GMIS}. In addition to that, the GMIS product is accompanied by a global Human Built-up and Settlement Extent (HBASE) data set \cite{HBASE,hotex}, which is used as a mask for GMIS. Both products are based on the 2010 Global Land Survey (GLS) Landsat data \cite{gutman2013assessment}. Here we used both of them as ancillary data sets for this study. The GMIS product was an input for the feature derivation steps, which will be discussed later, while the HBASE data set was used as an urban extent mask. Based on the HBASE product, all non-HBASE pixels were excluded from the prediction of BH/BV. Note that the HBASE product includes road networks from rasterized vector data. Because this study focuses on BH and BV, we excluded the road pixels in the HBASE product from subsequent analyses as well.

\subsection*{Training Data Derivation}
Based on the 1 m lidar-derived BH, BV for a 30 m pixels was estimated as:
\begin{eqnarray}
\label{eq:vol}
	BV = \sum_{i=1}^{N} bh_i\times A = A \times\sum_{i=1}^{N} bh_i,
\end{eqnarray}
where $N$ is the number of 1 m building pixels within the 30 m pixel, $bh_1$\char`\~$bh_N$ are the above ground heights of the 1 m building pixels and $A$ is the area of a 1 m pixel. Consequently, the average BH for the target 30 m pixel could be defined as:
\begin{eqnarray}
\label{eq:avgh}
	\bar{BH} = \frac{BV}{ \sum_{i=1}^{N} A} = \frac{1}{N} \sum_{i=1}^{N} bh_i.
\end{eqnarray}

Note that only 1 m building pixels within 30 m pixels were used to calculate $\bar{BH}$, which represents the average building height at 30 m resolution. For the sake of convenience, it is referred to as BH hereafter.

\subsection*{Feature Derivation}
Using the training data derived above, machine learning regression models based on features derived from Landsat and DSMs were trained to predict BH/BV. A total of 88 feature variables was used as inputs for the machine learning algorithms. They include pixel-level metrics, which were calculated on a pixel-by-pixel basis, and object-level metrics, which were calculated based on multi-level Landsat image objects. The following subsections describe how these features were calculated in detail.

\subsubsection*{Pixel-level Metrics}
The first set of pixel-level metrics used as input features are height values ($Z$) from the three DSMs described earlier. In addition, we also used slope calculated from the DSMs. Conceptually, it is notable that slope is also an important factor to consider. For example, slope information is key to many morphology-based filters for separating the ground height from surface height and the filtered results are generally less reliable in areas with steep terrains if the slope information is not properly considered \cite{geiss2015normalization,maguya2013adaptive,meng2010ground,zhang2003progressive}. Therefore, for each of the DSM datasets used, we also included its slope information using a calculation method proposed by \cite{horn1981hill}:
\begin{eqnarray}
\label{eq:sx}
	S_{i,j}^{x} = [(Z_{i+1,j+1}+2Z_{i+1,j}+Z_{i+1,j-1})-(Z_{i-1,j+1}+2Z_{i-1,j}+Z_{i-1,j-1})]/(8\Delta X),
\end{eqnarray}

\begin{eqnarray}
\label{eq:sy}
	S_{i,j}^{y} = [(Z_{i+1,j+1}+2Z_{i,j+1}+Z_{i-1,j+1})-(Z_{i+1,j-1}+2Z_{i,j-1}+Z_{i-1,j-1})]/(8\Delta Y),
\end{eqnarray}

\begin{eqnarray}
\label{eq:sy}
	S_{i,j} = \sqrt{(S_{i,j}^{x})^2+(S_{i,j}^{y})^2},
\end{eqnarray}
where $Z_{i,j}$ is the DSM height value of the pixel $(i,j)$; $\Delta X$ and $\Delta Y$ are the distances between pixels in the X and Y directions, respectively (30 m in this case); $S_{i,j}^{x}$ and $S_{i,j}^{y}$ are the slopes in the X and Y directions, respectively; and $S_{i,j}$ is the estimated slope of the pixel $(i,j)$.

Finally, IS fraction from the GMIS project was used as another pixel-level metric to incorporate urban density information into the modeling of BH/BV.

\subsubsection*{Object-based Metrics}
Image segmentation is a necessary preprocessing step for object-based feature derivation. The Landsat images used in this study were segmented using the Recursive Hierarchical Image Segmentation (RHSeg) software package \cite{tilton2012best}, a recursive approximation of the Hierarchical Image Segmentation (HSeg) algorithm. Thanks to its divide-and-conquer approach, RHSeg greatly improves the efficiency of the HSeg algorithm on cluster- or cloud-computing systems. The output of the RHSeg algorithm includes image objects at multiple levels of detail, where finer-level objects are nested within coarser-level objects. Segmentation level selection is needed to extract information efficiently from complicated hierarchical segmentation (20--80 levels for most Landsat images, depending on the image complexity). Based on heuristic analyses of the RHSeg results, we used segmentation results at three representative levels for feature derivation. The objects at these three levels were determined by three levels of the object size thresholds: 100 pixels (9 ha), 1,000 pixels (90 ha), and 10,000 pixels (900 ha), which represent roughly the scales of face-blocks, residential neighborhoods, and communities, respectively. This hierarchy of spatial scales is based on the definition of different urban spatial units in \cite{apa2006}.

Four groups of object-based metrics were derived using the Landsat-based segmentation as spatial units of feature calculation. The first group of object-based features included the mean ($\bar{Z}$), maximum ($Z_{max}$), minimum ($Z_{min}$), and standard deviation ($Z_{std}$) of the DSM height. These four variables were calculated for all three DSMs from objects at at all three segmentation levels. The rationale for including these features was that the maximum and minimum of the height are related to the roof-top and ground heights within an object, while the mean and standard deviation describe the general pattern of the height distribution in an object.

The second group of features included the mean ($\bar{IS}$), maximum ($IS_{max}$), and minimum ($IS_{min}$) of IS values within the objects. These features were included to help separating areas with different urban densities. On a conceptual level, when deriving BH from DSM data, different ground and roof-top height estimation strategies should be adopted in areas with different urban densities. For example, for low density urban areas, minimum DSM height may be a good estimation of ground height, while minimum DSM height may overestimate the ground height in dense urban areas because of the absence of pure ground pixels. Therefore, these 2D urban information might be useful for the BH/BV estimation as well.

Combining DSM heights and IS fractions, we derived the third group of features. First, for an object $O$, the IS-weighted mean of height was calculated as:

\begin{eqnarray}
\label{eq:zis}
	\bar{Z}_{IS} = \frac{\sum_{(i,j) \in O} Z_{i,j} IS_{i,j}}{\sum_{(i,j) \in O} IS_{i,j}},
\end{eqnarray}
where $\bar{Z}_{i,j}$ and $\bar{IS}_{i,j}$ were the height value and IS fraction of a pixel within the object $O$, respectively. The rationale for including this feature was using IS to separate the height of non-building vertical structures such as trees from BH at the sub-pixel level. 

Similarly, two other features based on the integration of DSM and IS were also included:
\begin{eqnarray}
\label{eq:zismin}
	\bar{Z}_{IS_{min}} = \frac{\sum_{(i,j) \in O_0} Z_{i,j}}{|O_0|},
\end{eqnarray}
\begin{eqnarray}
\label{eq:zismax}
	\bar{Z}_{IS_{max}} = \frac{\sum_{(i,j) \in O_1} Z_{i,j}}{|O_1|},
\end{eqnarray}
where $O_0$ and $O_1$ were subsets of the object $O$ where IS took minimum and maximum values, respectively.

Finally, the fourth group included features characterizing the average slope of objects. For an object $O$, the average slope $S_O$ was estimated by calculating the average slopes in the $X$ ($S_O^x$) and $Y$ ($S_O^y$) directions first:

\begin{eqnarray}
\label{eq:sox}
	S_O^x = \frac{1}{|O|}\sum_{(i,j) \in O} S_{i,j}^x,
\end{eqnarray}

\begin{eqnarray}
\label{eq:soy}
	S_O^y = \frac{1}{|O|}\sum_{(i,j) \in O} S_{i,j}^y,
\end{eqnarray}

\begin{eqnarray}
\label{eq:so}
	S_O =  \sqrt{(S_{O}^{x})^2+(S_{O}^{y})^2}.
\end{eqnarray}

Table~\ref{table2} lists all features described above. For a given 30 m pixel, features derived at the pixel-level, and from level-1, level-2, and level-3 objects containing the pixel were stacked together to form the complete feature set. 

\begin{table}[!ht]
\begin{adjustwidth}{-2.25in}{0in} 
\centering
\caption{
{\bf List of input features used for machine learning regression models including features derived from AW3D30, GDEM, SRTM, and Landsat-based impervious surface (IS). The right column shows the count of features at different levels.}}
\begin{tabular}{cccccc}
  \toprule
  Level &\multicolumn{4}{c}{Data Source} & Feature Count\\
 \midrule
\midrule 
\addlinespace \\
  & AW3D30 & SRTM & GDEM & IS & \\
   \cmidrule{2-5}
\addlinespace \\
  \multirow{1}{*}{Pixel} & $Z_{i,j}$, $S_{i,j}$ & * & * & $IS$ & 7 \\
  \midrule
  \multirow{4}{*}{}Level-1 object & $\bar{Z}$, $Z_{max}$, $Z_{min}$, $Z_{std}$, $\bar{Z}_{IS}$, $\bar{Z}_{{IS}_{max}}$, $\bar{Z}_{{IS}_{min}}$, $S_O$ & * & * & $\bar{IS}$, $IS_{max}$, $IS_{min}$ & 27 \\
  Level-2 object & Same as Level-1 & * & * & Same as Level-1  & 27 \\
Level-3 object & Same as Level-1 & * & * & Same as Level-1  & 27 \\
  \midrule
  \multirow{1}{*}{Stacked}& -  & - & -  & - & 88 \\
  \bottomrule
\end{tabular}
\vspace{1ex}
\raggedright \textsuperscript{*}The features derived from SRTM and GDEM are the same as those derived from AW3D30.
\label{table2}
\end{adjustwidth}
\end{table}

\subsection*{Machine Learning Experiments}
Using features derived in the previous section and the lidar-based training data, we tested different machine learning regression techniques to estimate BH/BV. Note that although most of the features were derived at the object level, the machine learning modeling and prediction was performed at the pixel level because pixel-level feature concatenation was performed.

Two sets of experiments were designed to test (1) the usefulness of input features derived from different DSMs, and (2) the efficacy of different machine learning algorithms in predicting BH/BV. For the first group of experiments, we performed random forest (RF; \cite{breiman2001random}) regression of the BH based on features derived from AW3D30, GDEM, SRTM only, and the combination of all features. The IS-based features were used in all four tests. For the second experiment, two most widely used regression algorithms in remote sensing, RF and Cubist™ \cite{breiman2001random,rulequest2016}, were selected for BH/BV estimation. The results of these two experiments were used as a basis for selecting features and the machine learning algorithm for producing the final BH/BV maps.

\subsection*{Validation of Estimated BH/BV}
The validation of BH/BV is greatly limited by the availability of reference data. In this study, the lidar-based training data was the most accurate reference data available. Therefore, to interpret the machine learning experiments described above, a 10-fold cross-validation (CV) was adopted, which was also used to validate the final BH/BV maps. Instead of dividing the training data randomly into CV subsets, we divided the Landsat scenes into 10 random groups with roughly equal number of training pixels and then assigned the training pixels into subsets based on the scene they belong to. This scene-level cross-validation (SLCV) approach was designed to avoid inflated accuracy estimates due to spatial autocorrelation between training and testing samples when spatially adjacent pixels are divided into training and testing for CV \cite{friedl1999maximizing}. Using SLCV, we produced accuracy estimates, including the root-mean-square error (RMSE) and correlation coefficient (R\textsuperscript{2}). These accuracy estimates, as explained above, should be reasonably unbiased because spatial autocorrelations between training and testing have been minimized by the SLCV approach.

Additional validation of BV was done based on correlations between the estimated BV and socioeconomic variables. We used a data set comprised of socioeconomic variables (including population, gross domestic product (GDP), and transportation CO\textsubscript{2} emission) for 12 metropolitan areas defined by the Organization for Economic Co-operation and Development (OECD; \cite{brezzi2012redefining}). Population and GDP are two important socioeconomic variables for urban areas that have been found to be correlated with built-up area and building volume \cite{cheng2011trend,qiu2010spatial,avelar2009linking,national1998people,lu2011volumetric,lu2010population}. The transportation CO\textsubscript{2} emission data was used here as a proxy of energy consumption from the transportation sector, which is also impacted by the 2D and 3D configurations of urban areas \cite{resch2016impact}. We did not include the emission from other sectors, which might be more impacted by other exogenous factors such as local climate \cite{baiocchi2015spatial}. Regression analyses were performed between these variables and estimated BV. Based on existing studies on the relationship between BV and these variables, we considered strong correlations as indicators of a good performance for the BV mapping method.

\section*{Results}
\subsection*{Comparison of Input Features Based on Different DSMs}
\begin{figure}[!h]
\begin{adjustwidth}{-2.25in}{0in} 
\caption{{\bf Scatterplot-based comparison of the accuracies of the building height using random forest and input features from different elevation datasets.} The points in the scatterplots were derived from the 10-fold cross-validation.}
\label{fig3}
\includegraphics[width=7.5in]{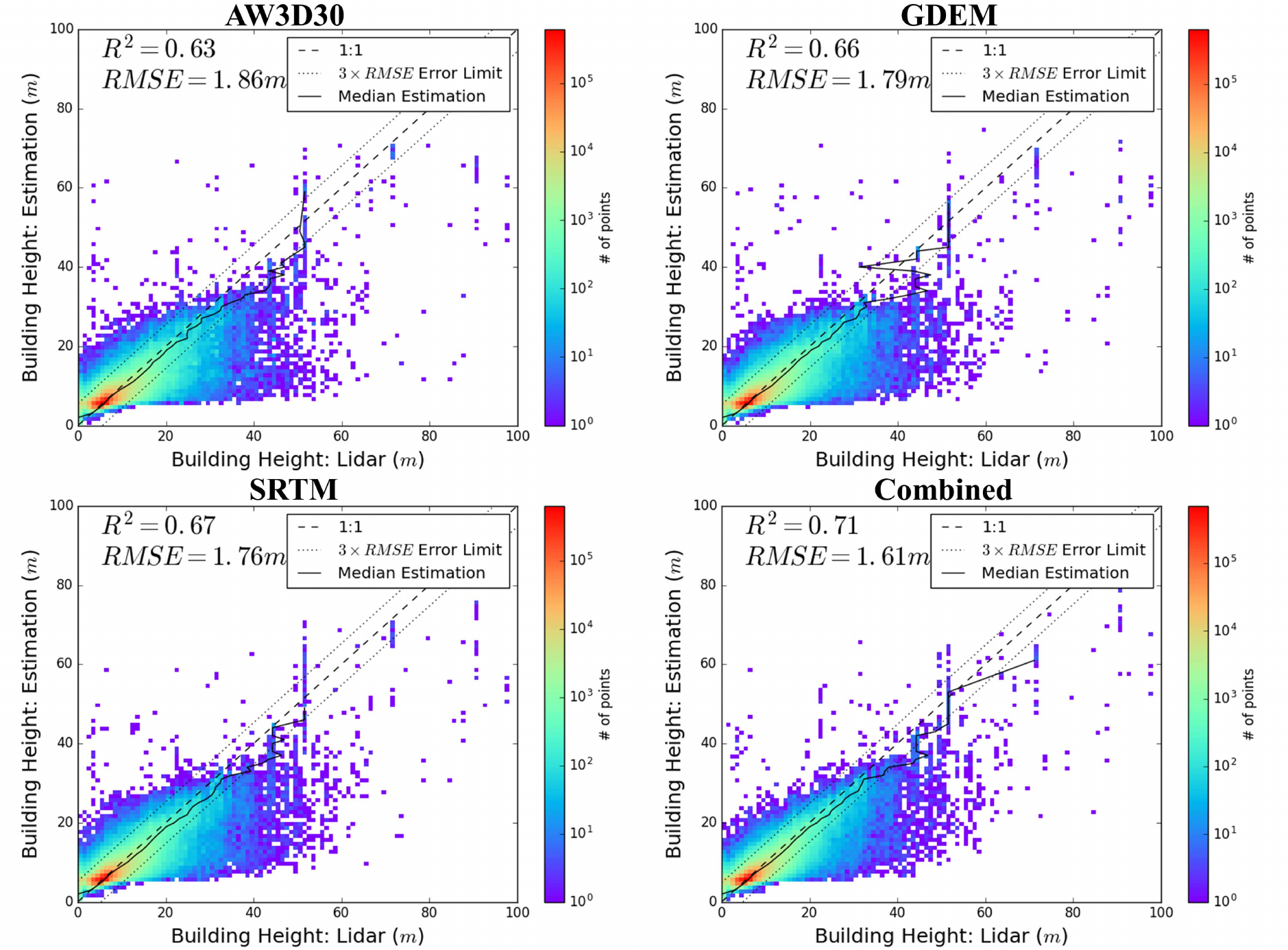}
\end{adjustwidth}
\end{figure}

Our comparisons of the effectiveness of different features focused on the height estimation accuracy using RF. Surprisingly, AW3D30 produced the worst result in terms of both RMSE and R\textsuperscript{2} (Fig \ref{fig3}). The SRTM-based result was slightly better than the result using ASTER GDEM, while the best result was achieved using input features derived from all three elevation datasets together. These results confirmed the hypothesis that the DSMs complement each other. Thus, the final BH and BV maps will be produced using the combination of all the features.

\begin{figure}[!h]
\begin{adjustwidth}{-2.25in}{0in} 
\caption{{\bf The spatial distribution of (a) quality flag of the AW3D30 elevation data set, where black and blue mark no-data and low data-reliability areas, respectively, (b) error of height estimation using random forest, and (c) error of height estimation using the Cubist™ regression tree.} The zoom-in window shows the distribution of height errors for the greater London area.}
\label{fig4}
\includegraphics[width=7.5in]{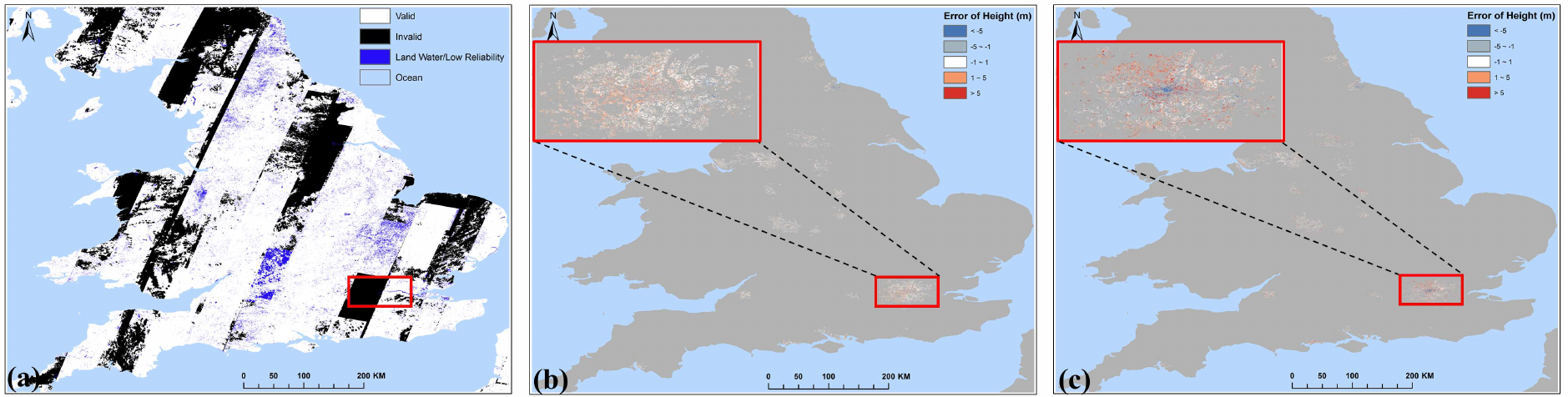}
\end{adjustwidth}
\end{figure}

The relatively higher RMSE (1.86 m) of BH using AW3D30 compared to using ASTER GDEM (1.79 m) and SRTM (1.76 m) contradicts the fact that AW3D30 is expected to have a higher vertical accuracy, particularly in urban areas. This is probably driven by the relatively high amount of data gaps in the AW3D30 product. Fig \ref{fig4} shows that a large portion of the study area lacks AW3D30 data or has low-reliability data. In these areas, the error of the BH was clearly higher than in areas with valid AW3D30 data, as shown by the zoom-in views of Fig \ref{fig4}, where a boundary is present between East and West London in terms of BH error, which coincided with the boundary of missing data. 

\subsection*{Comparison of Results using Random Forest and Cubist™}
\begin{figure}[!h]
\caption{{\bf The histogram of error in predicted height using (a) random forest and (b) Cubist™ regression tree.} The histograms for three groups of pixels where AW3D30 data is valid, invalid, and all pixels are plotted using different colors.}
\label{fig5}
\includegraphics[width=5.2in]{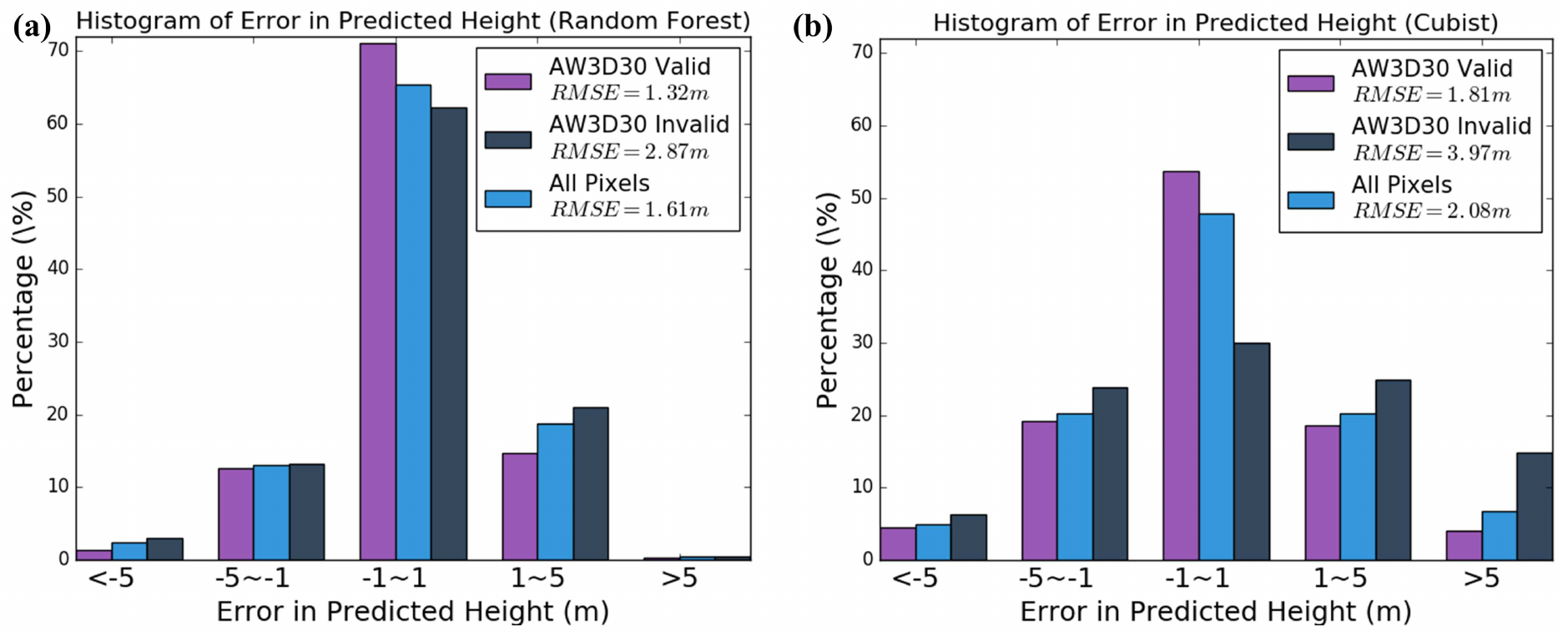}
\end{figure}

Based on the comparison of the spatial distribution of errors of RF and Cubist™ (see Fig \ref{fig4}), it is clear that missing data significantly affects how RF and Cubist™ perform. The likely cause is that these two machine learning algorithms employ different missing data handling mechanisms. When BH predictions were made on pixels containing missing feature variables, the RF software used in this study (a C++ implementation) made predictions based on trees that do not involve the missing features, while Cubist™ replaced the missing values with average values (derived from the entire training data) for each missing feature, which is a statistical technique known as imputation. The imputation-based method is clearly problematic in this case because the final prediction might be made based on the imputed values. Therefore, the Cubist™ results exhibited more difference in height estimation error between areas with and without valid AW3D30 data. To further investigate how missing data affected the performance of BH estimation, we examined the histograms of BH error for pixels with valid AW3D30 data, invalid AW3D30 data, and all pixels (see Fig \ref{fig5}). When valid AW3D30 data were available, BH estimation error was consistently lower than that of the pixels with invalid AW3D30 data and all pixels. In fact, the error histograms for all pixels fall in the middle of the other two, suggesting that AW3D30 could have achieved better BH estimation performance if much less data gaps were present in the study area. Again, the error histograms suggest that Cubist™ is more affected by missing data.

\begin{figure}[!h]
\begin{adjustwidth}{-2.25in}{0in} 
\caption{{\bf Scatterplot-based comparison of the accuracies of height and volume estimation using random forest (RF) and Cubist™ regression tree.} The points in the scatterplots were derived from the 10-fold cross-validation.}
\label{fig6}
\includegraphics[width=7.5in]{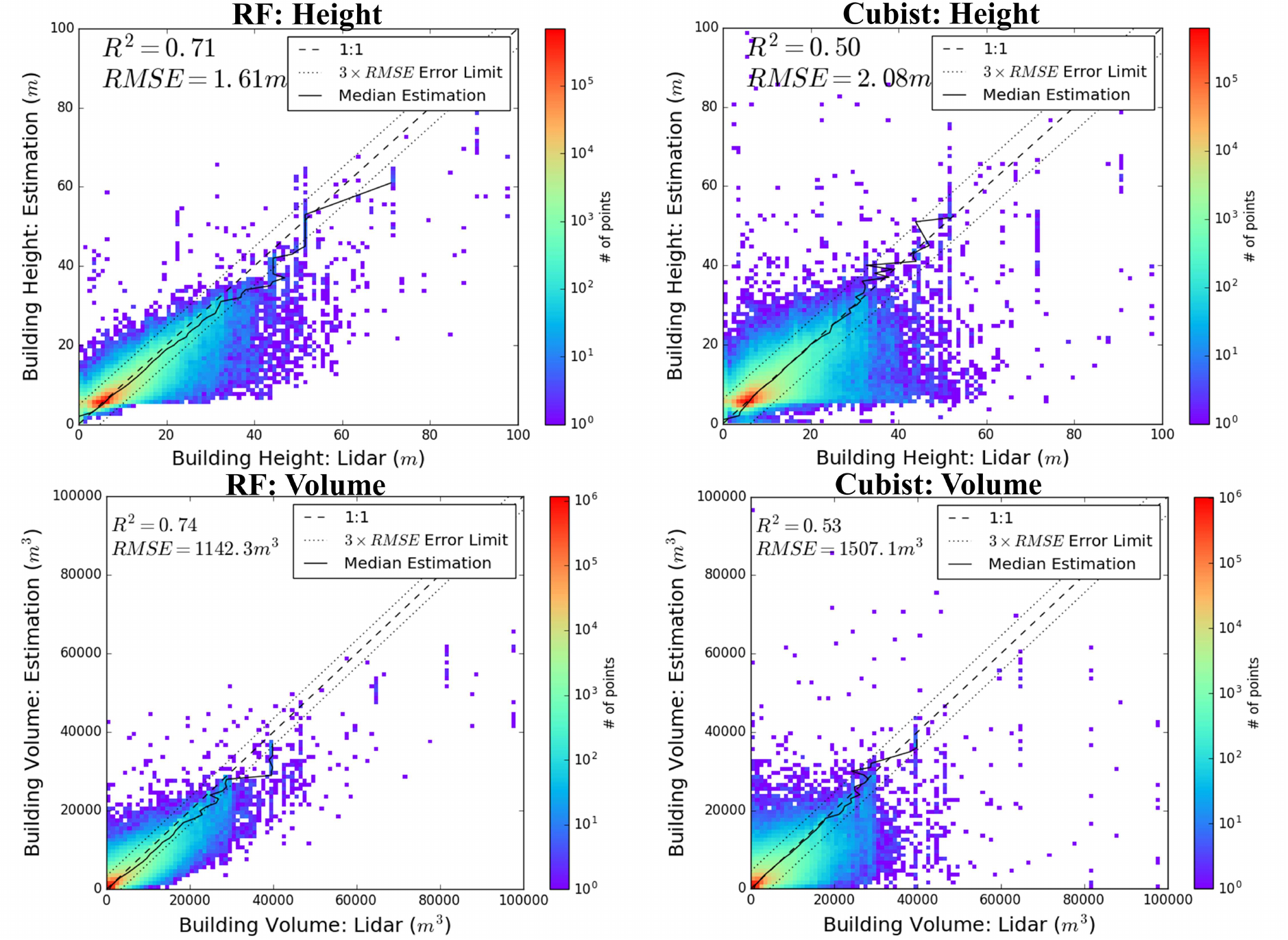}
\end{adjustwidth}
\end{figure}

The CV results also indicated that RF outperformed the Cubist™ regression tree in estimating both BH and BV (see Fig \ref{fig6}). Using RF, an RMSE of 1.61 m was achieved for BH estimation, which is 22\% lower than that of the Cubist™ regression tree. And an RMSE of 1,142 m\textsuperscript{3} was obtained for BV estimation using RF, which 24\% lower than that of the Cubist™ regression tree. Furthermore, BH/BV estimated using RF had significantly better correlation with lidar reference data. Therefore, RF was chosen to produce the final BH and BV maps. The CV scores for RF-based BH and BV represent our best estimates of the accuracy of the final maps.

\subsection*{Characteristics of the Mapped BH and BV}
\begin{figure}[!h]
\begin{adjustwidth}{-2.25in}{0in} 
\centering
\caption{\bf Maps of (a) building height and (b) building volume for England.}
\label{fig7}
\includegraphics{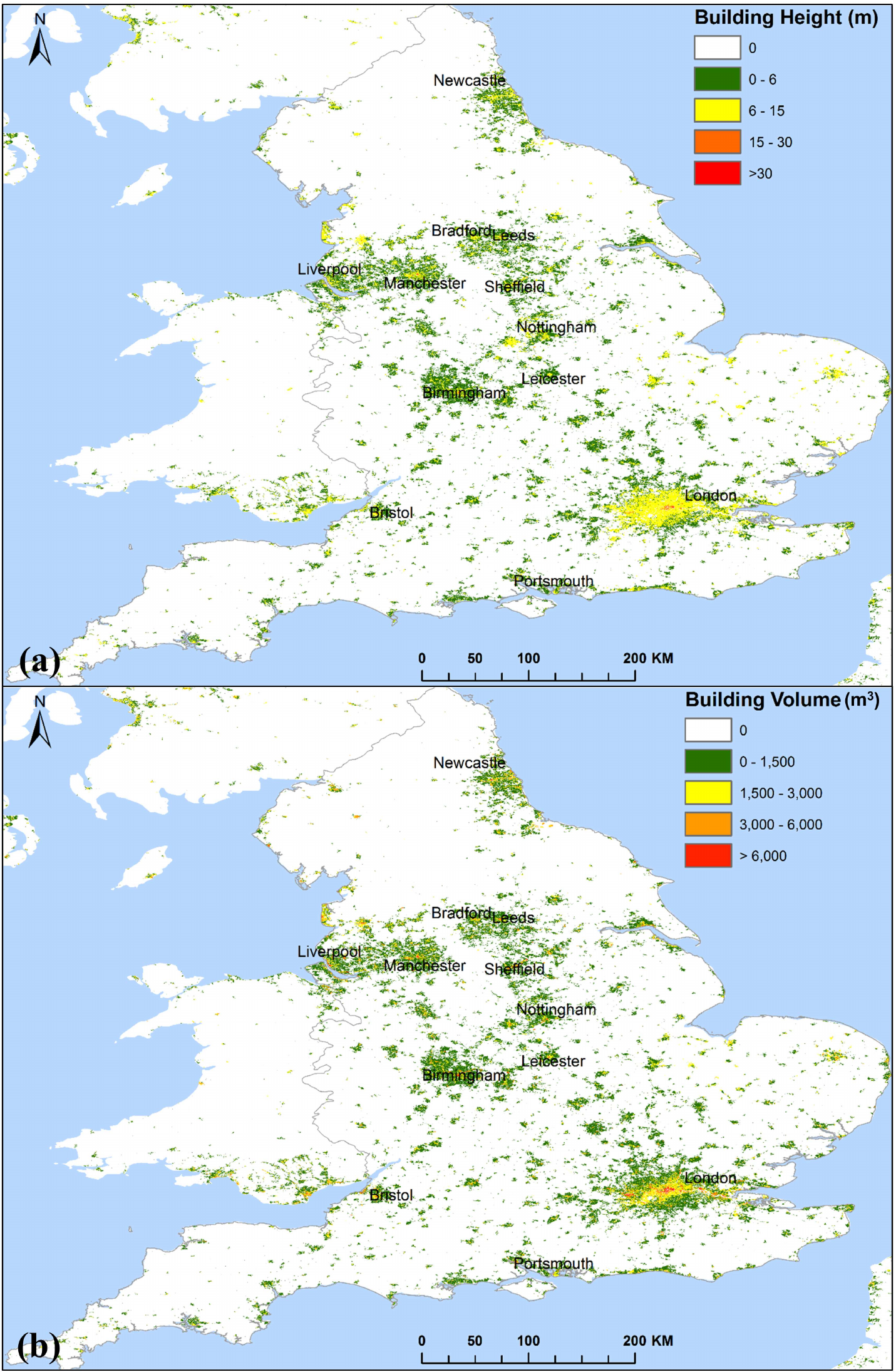}
\end{adjustwidth}
\end{figure}

Using the combination of all the derived features and the RF algorithm, the final BH/BV maps were produced. As shown in Fig~\ref{fig7}, the spatial distributions of the mapped BH and BV exhibited a strong pattern of agglomeration. The majority of urbanized areas in England had less than 6 m of BH and less than 1,500 m\textsuperscript{3} of BV. For the bulk of greater London and the core areas of several large cities (e.g., Birmingham and Manchester), however, the mapped BH was approximately 6 to 15 m and the BV was 1,500 to 3,000 m\textsuperscript{3}. Finally, BH higher than 30 m were found mostly distributed in the core area of the city of London and a few hub cities including Birmingham and Manchester.

\begin{figure}[!h]
\begin{adjustwidth}{-2.25in}{0in} 
\caption{{\bf 3D rendition of the building volume of England.}}
\label{fig8}
\includegraphics[width=7.5in]{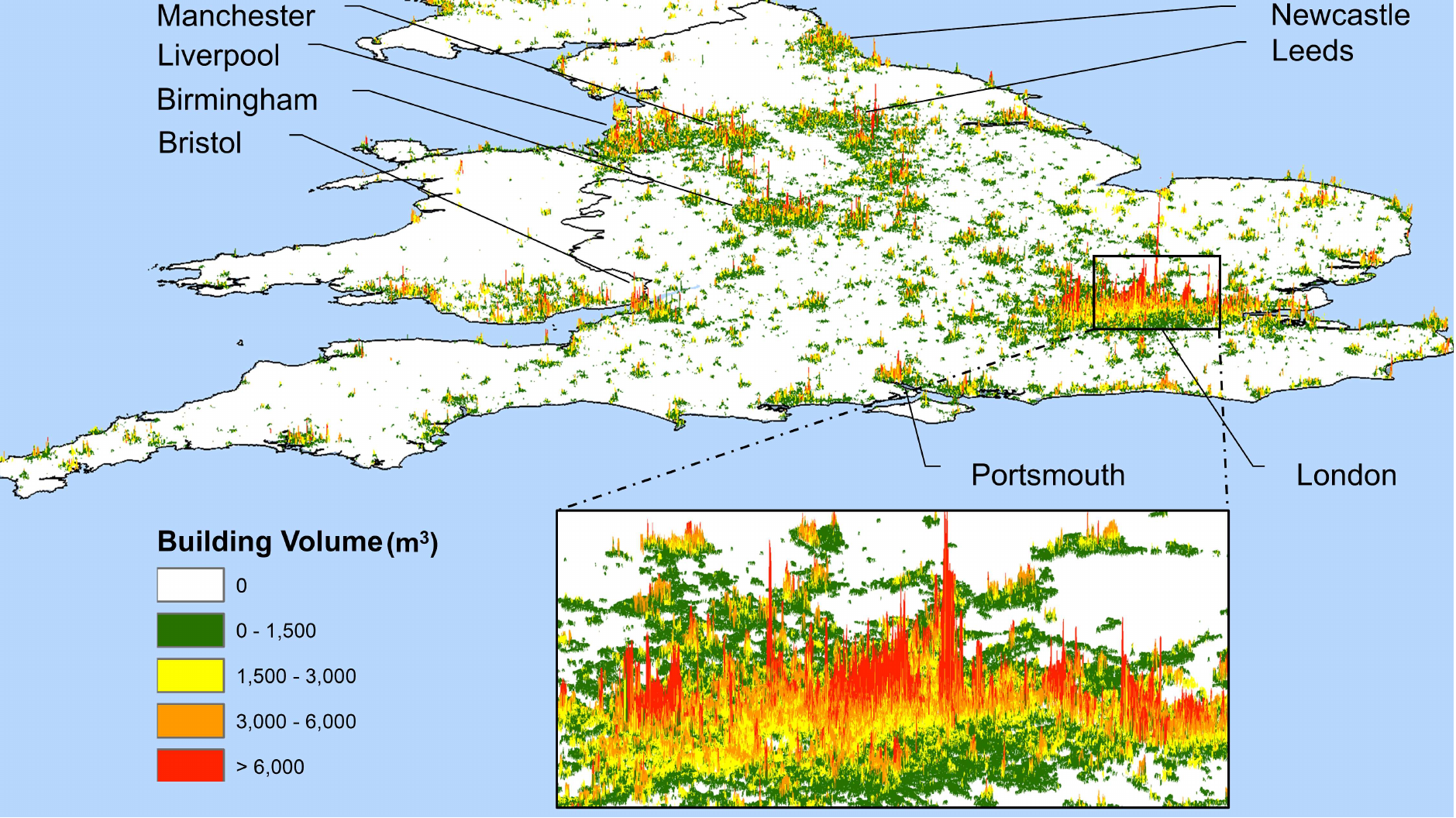}
\end{adjustwidth}
\end{figure}

We further examined the quality of the mapped BV using the 3D rendition function of the ArcScene software (see Fig~\ref{fig8}). The mapped BV captured well both the spatial distribution of urban areas in England and the concentration of BV within core urban areas. As shown by the zoom-in view of London, our method was able to reproduce a great amount of spatial details with medium resolution input data.

\begin{figure}[!h]
\begin{adjustwidth}{-2.25in}{0in} 
\caption{ {\bf Mapped building height and volume for the city of London, in comparison with lidar derived reference data (gray areas do not have lidar coverage).} The error of height and volume are derived from the difference between random forest predictions and lidar reference data.}
\label{fig9}
\includegraphics[width=7.5in]{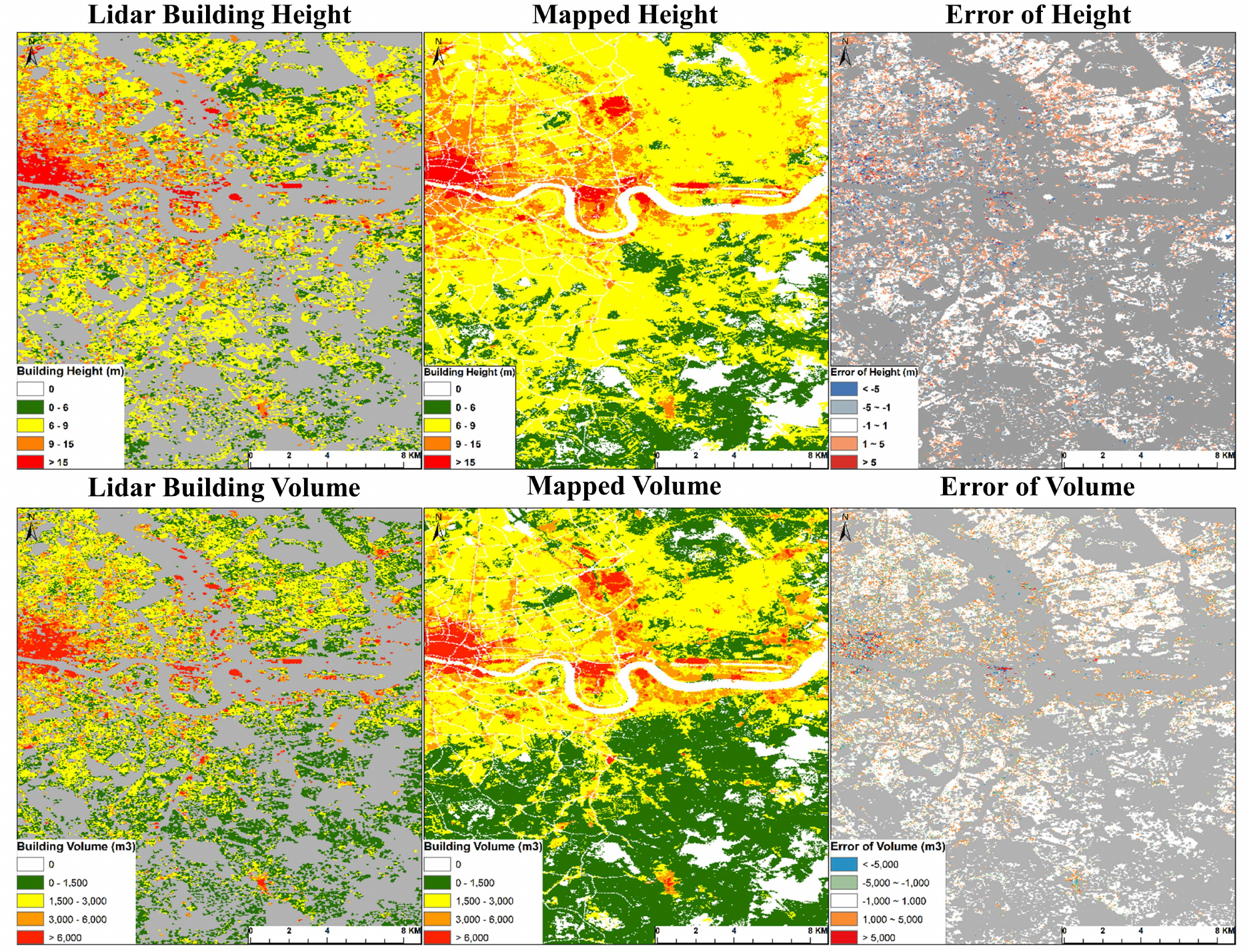}
\end{adjustwidth}
\end{figure}

More detailed maps and side-by-side comparisons with lidar data for the city of London are shown in Figures~\ref{fig9}. The predicted BH and BV matched the spatial patterns of lidar data very well. However, the prediction errors tended to be higher in urban centers because the used DSMs do not have a sufficient spatial resolution to map the complex height variations in the urban centers effectively.

\begin{figure}[!h]
\begin{adjustwidth}{-2.25in}{0in} 
\caption{{\bf The scatterplots between (a) total Human Built-up And Settlement Extent (HBASE) area and population, (b) total impervious surface and population, (c) total building volume and population, (d) total HBASE area and GDP, (e) total impervious surface and GDP, (f) total building volume and GDP, (g) total HBASE area and transportation CO\textsubscript{2} emission, (h) total impervious surface and transportation CO\textsubscript{2} emission, and (i) total building volume and transportation CO\textsubscript{2} emission.} Each point represents a metropolitan area defined by the Organization for Economic Co-operation and Development (OECD). The blue and green fitted lines and corresponding correlation coefficients are derived with London included and excluded, respectivelys.}
\label{fig10}
\includegraphics[width=7.5in]{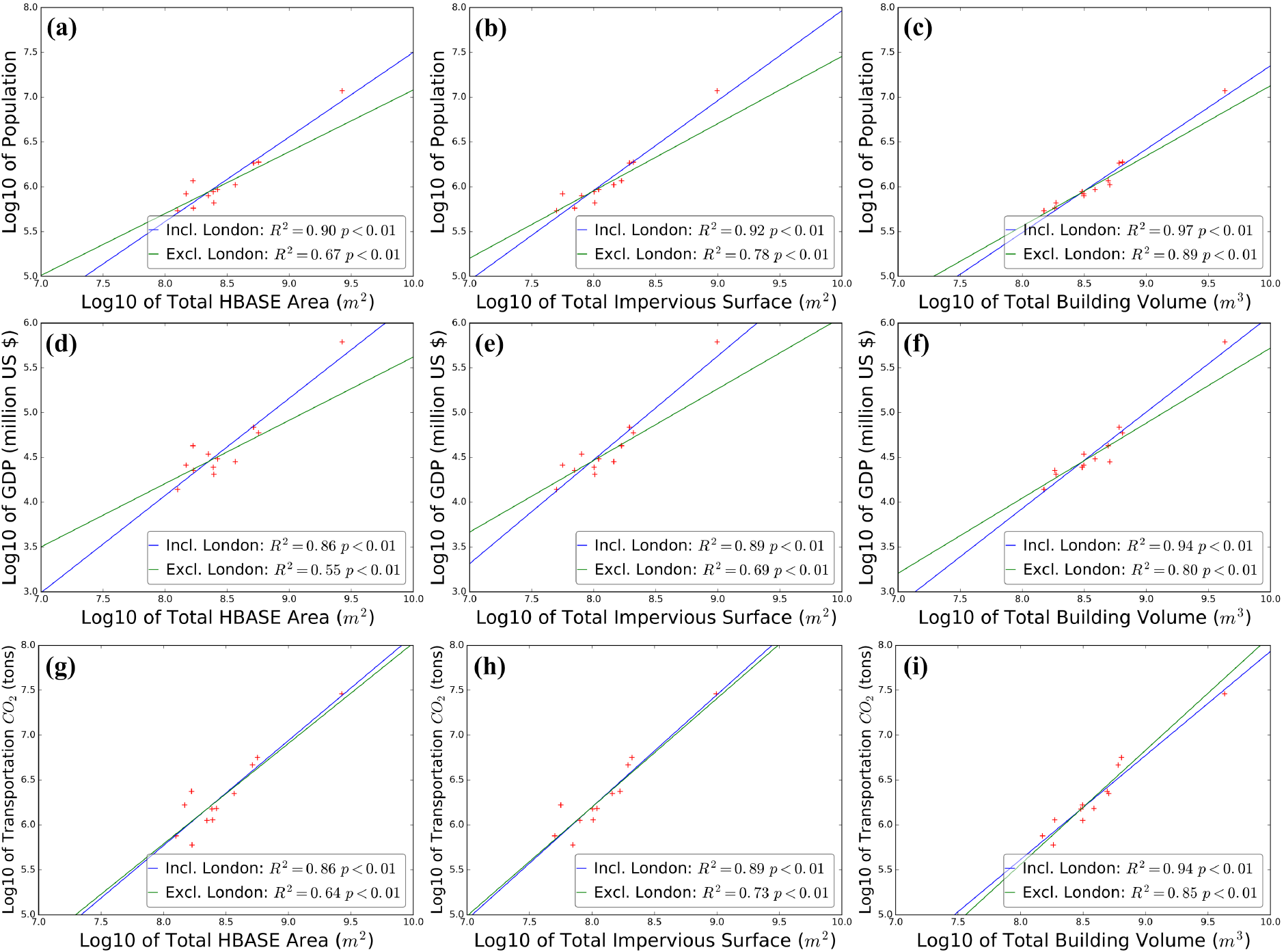}
\end{adjustwidth}
\end{figure}

To assess the overall quality of the mapped BV in terms of correlation with socioeconomic variables, statistical analyses were performed for 12 metropolitan areas defined by the OECD. Table \ref{table3} lists the total HBASE area, IS, and BV of these metropolises, alongside population, gross domestic product (GDP), and transportation CO\textsubscript{2} emission data from the OECD metropolitan database. As shown in Fig~\ref{fig10}, the correlation coefficients (R\textsuperscript{2}) between total BV and population, GDP, and transportation CO\textsubscript{2} emission was 0.97, 0.94, and 0.94 (0.89, 0.80, and 0.85 excluding London), respectively (on a logarithmic scale). Apparently,  BV is highly correlated with population, GDP, and transportation CO\textsubscript{2} emission. However, caution must be taken when interpreting the results of the correlation analyses. Since the BV-population relationship could be driving the correlation between BV and other socioeconomic variables, the high correlation scores do not necessarily indicate the predictive power of building volume for socioeconomic conditions. Even the BV-population relationship may not hold given a larger geographic area with more heterogeneous socioeconomic conditions. Also, because London is an order of magnitude larger than other cities, we also calculated correlation coefficients without London to avoid biased correlation coefficients. Nevertheless, Fig~\ref{fig10} does provide strong evidence that the mapped BV has satisfactory accuracy. In addition, we compared  the correlations between HBASE/IS and the socioeconomic variables to those of BV. According to Fig~\ref{fig10}, BV explained the variance of the socioeconomic variables significantly better than than HBASE and IS, suggesting that BV is useful for various socioeconomic studies and that BV has an obvious advantage over traditional 2D LCLU variables in such studies.

\begin{table}[!ht]
\begin{adjustwidth}{-2.25in}{0in} 
\centering
\caption{
{\bf Total Human Built-up and Settlement Extent (HBASE) area, impervious surface (IS) area, and building volume (BV) of 12 metropolitan areas in England defined by OECD.} Population, gross domestic product (GDP), and transportation CO\textsubscript{2} emission data from the Organization for Economic Co-operation and Development (OECD) metropolitan database are listed alongside.}
\begin{tabular}{C{2cm} C{2.5cm} C{2.5cm} C{2.5cm} C{2.5cm} C{2cm} C{2cm}}
  \toprule
  Name & Population & GDP (bil. US \$) & Transportation CO\textsubscript{2} (mt) & HBASE (km\textsuperscript{2}) & IS (km\textsuperscript{2}) & BV (km\textsuperscript{3})\\
\midrule
\midrule \\
London&11,793,530&616.07&28.89&2671.22&986.80&4.30\\
\midrule \\
Birmingham&1,884,199&59.46&5.63&565.44&208.78&0.64\\
\midrule \\
Manchester&1,841,382&68.67&4.66&494.70&167.63&0.49\\
\midrule \\
Leeds&1,166,267&42.64&2.37&125.84&50.29&0.15\\
\midrule \\
Newcastle&1,050,561&28.33&2.23&264.69&109.64&0.38\\
\midrule \\
Liverpool&929,014&30.43&1.52&515.29&194.49&0.61\\
\midrule \\
Sheffield&880,237&24.54&1.51&244.32&100.77&0.30\\
\midrule \\
Nottingham&835,625&25.92&1.66&222.60&80.05&0.31\\
\midrule \\
Bristol&795,481&34.53&1.12&368.00&145.25&0.51\\
\midrule \\
Leicester&660,817&20.45&1.14&247.04&102.18&0.31\\
\midrule \\
Portsmouth&577,191&22.66&0.60&168.89&70.27&0.19\\
\midrule \\
Bradford&540,172&13.87&0.76&147.40&56.17&0.18\\
\bottomrule
\end{tabular}
\label{table3}
\end{adjustwidth}
\end{table}

\section*{Discussion}
The machine learning regression experiments clearly demonstrated that global DSMs have the potential for mapping large-scale urban structures. However, several limitations exist in this study and should be addressed in future studies. First, due to inconsistent acquisition time and cloud cover, temporal discrepancies exist between passive optical (Landsat in this case), DSM, and lidar data sets. Such discrepancies could introduce great uncertainties when there are mismatches between (a) features extracted from different data sources and/or (b) training data and features. In this study, the impacts of these discrepancies have been minimized by choosing England as the study area. The relatively slow change in England enabled this study to accurately map BH/BV using data from a wide temporal range. However, for other regions in the world with higher rates of urbanization, temporal discrepancy cloud be a much greater issue. Possible solutions include using recent elevation data sets (e.g, TanDEM-X DEM \cite{zink2014tandem}), which may have better agreement with lidar and passive optical data in acquisition time.

Another limitation of this study is the relatively higher error in urban center areas as a result of insufficient spatial resolution to resolve complex height variations in such areas. Again, this is, to a large degree, a data availability issue. Higher resolution DSM data, such as the TanDEM-X DEM (12 m \cite{zink2014tandem}), are promising for more accurate height measurements in dense urban areas \cite{geiss2015normalization}. However, more studies are needed to determine is there is any added value from using these higher resolution DSMs under the proposed methodological framework.

It is not clear to what extent the Landsat-derived urban extent map and other Landsat-based inputs (IS map and segmentation results) affected the mapping accuracy. Because Landsat-based segmentation was used as spatial units of height information extraction, inaccurate delineation of building patches may introduce biases and errors in estimated height. It would be worthwhile to explore building patch delineation with both passive optical and DSM information in future studies. Also, more systematic investigations are needed to determine the optimal strategy to fuse passive optical and DSM data for BH/BV mapping under different terrain and urban density conditions. Finally,  the Sentinel-2 mission has been generating multi-spectral imagery with 10--20 m spatial resolution \cite{drusch2012sentinel}. With higher spatial resolution, the Sentinel-2 imagery could be a more suitable data source for our optical-DSM fusion method, particularly for fusion with higher resolution DSMs.

Besides methodological refinements, linking BV and socioeconomic variables could be another promising future research direction. The linkages between earth observation information with demographic and socioeconomic data has been of great interest \cite{national1998people} and have been exploited to provide vital information when traditional data source is not available. Nighttime lights data have been proven a valuable source of information for large-scale economic activity, cultural patterns, and conflicts \cite{henderson2012measuring,roman2015holidays,li2014can}. LCLU information has been an integral part of the effort of using census data to produce gridded population data products \cite{dobson2000landscan,lloyd2017high}. In our analyses of 12 England metropolitan areas, it was clear that IS was better linked with socioeconomic variables and that BV was better than both HBASE and IS. This was not surprising because, by using HBASE, IS, and BV, progressively more information was provided. With the addition of 3D information, BV has an advantage over 2D LCLU variables for the integration of remote sensing and social science and has the potential of breaking new grounds in this field. Future studies will examine if such advantage remains given a larger area and a larger set of socioeconomic variables.

\section*{Conclusion}
Urbanization is a 3D phenomenon that needs to be observed through remote sensing in not only the 2D space but also the vertical domain. We presented an innovative method combining the strength of Landsat imagery and freely available global elevation data sets to spatially map BH/BV across all of England. It was demonstrated that the proposed method can achieve a reasonably high accuracy, even with sub-optimal data. The method was applied to England to produce a freely available BH/BV data set for the entire country \cite{EnglandUrban3D}. Based on cross-validation using the training data derived from lidar measurements, the RMSE of BH was only 1.61 m. And the RMSE of BV was 1,142.3 m\textsuperscript{3}. Also, the mapped BV had strong correlation with population, GDP, and transportation CO\textsubscript{2} emission variables, indicating a good overall performance of the BV mapping method and an advantage of using the mapped BV in socioeconomic studies over traditional 2D LCLU data.

Despite great  data availability limitations, the proposed method exhibits great potential for large-area characterization of urban structures. Spatial maps of urban structural information, including BH and BV, are important for many urban applications. With the increasing availability of global elevation data, our method may provide a path towards global products of 3D urban structures, which will greatly enhance our understanding of urbanization, particularly in the fields of urban vulnerability to natural disasters, urban heat island effect, and sustainable urbanization. 

\section*{Acknowledgments}
This study was made possible by the support of the Dr. Jingli Yang Geographical Sciences Research Fellowship. We thank the U.S. Geological Survey (USGS) for providing the Landsat surface reflectance data product and Emu Analytics for making open data on building height of England available.


%

\end{document}